\documentclass[twocolumn,preprintnumbers,nofootinbib,amsmath,amssymb,amsfonts,aps,prd,balancelastpage,superscriptaddress]{revtex4-2}

\usepackage{color}
\usepackage{mathtools}
\usepackage{amsthm,amstext,amscd}
\usepackage{epsfig,graphicx,bm}
\usepackage{epstopdf, epsf}
\usepackage{dcolumn}
\usepackage{tensor}

\usepackage{datetime}
\newdateformat{monthyeardate}{%lll
  \monthname[\THEMONTH], \THEYEAR}
  
\usepackage{graphicx,caption,subcaption}
\usepackage[colorlinks,
           linkcolor = blue,		
           citecolor = blue,
           ]{hyperref}

\newcommand{\be}{\begin{equation}}
\newcommand{\ee}{\end{equation}}

\newcommand{\bse}{\begin{subequations}}
\newcommand{\ese}{\end{subequations}}
\newcommand{\bea}{\begin{eqnarray}}
\newcommand{\eea}{\end{eqnarray}}
\newcommand{\ba}{\begin{array}}
\newcommand{\ea}{\end{array}}
\newcommand{\bc}{\begin{center}}
\newcommand{\ec}{\end{center}}

\def\ca{{\cal A}}
\def\cb{{\cal B}}

\def\cg{{\cal G}}

\def\co{{\cal O}}

\def\ba{{\bf A}}

\def\bc{{\bf C}}

\def\be{{\bf E}}

\begin{document}

\vspace*{3mm}

\title{(Lovelock)$^2$ inflation: explaining the ACT data and equivalence to Higgs--Gauss--Bonnet inflation}

\author{Andrea Addazi}
\email{andrea.addazi@qq.com}
\affiliation{School of Physics and Astronomy, Anqing Normal University, Anqing 246133, People's Republic of China}
\affiliation{Institute of Astronomy and Astrophysics, School of Mathematics and Physics, Anqing Normal University, Anqing 246133, China}
\affiliation{CAS Key Laboratory for Research in Galaxies and Cosmology, School of Astronomy and Space Science, University of Science and Technology of China, Hefei 230026, China}
\affiliation{Laboratori Nazionali di Frascati INFN, Frascati (Rome), Italy, EU}

\author{Yermek Aldabergenov}
\email{ayermek@fudan.edu.cn}
\affiliation{Department of Physics, Fudan University, 220 Handan Road, Shanghai 200433, China}

\author{Daulet Berkimbayev}
\affiliation{Department of Theoretical and Nuclear Physics, Al-Farabi Kazakh National University, 71 Al-Farabi Ave., Almaty 050040, Kazakhstan}

\author{Yifu Cai}
\email{yifucai@ustc.edu.cn}
\affiliation{Deep Space Exploration Laboratory/School of Physical Sciences,
University of Science and Technology of China, Hefei, Anhui 230026, China}
\affiliation{CAS Key Laboratory for Researches in Galaxies and Cosmology/Department of Astronomy,
School of Astronomy and Space Science, University of Science and Technology of China, Hefei, Anhui 230026, China}

\date{\monthyeardate\today}

\begin{abstract}
\noindent

We revisit the Starobinsky model of inflation in light of recent data from the Atacama Cosmology Telescope (ACT), which indicates a potential preference for a slightly larger scalar spectral index $n_s$ than predicted by the standard $R^2$ scenario. We demonstrate that a natural one-parameter generalization to a quadratic model $\sim L+L^2$ in the Lovelock invariant $L=R+\frac{\alpha}{4}\cg$ ($\cg$ is the Gauss--Bonnet term), can effectively resolve this minor tension. Scalar-tensor formulation of this theory yields an Einstein-frame Starobinsky-type scalar potential augmented by Gauss--Bonnet and derivative couplings, which modify the inflationary slow-roll dynamics. We show that a non-zero coupling $\alpha$ for the Gauss-Bonnet term can shift $(n_s, r)$ along a trajectory that brings the predictions into better agreement with the ACT likelihood. We also find that $L+L^2$ gravity, in its scalar-tensor formulation, is equivalent to Higgs inflation coupled to the Gauss--Bonnet term, and belongs to the Horndeski/galileon class of modified gravities. This work establishes the quadratic $f(L)$ gravity as a compelling and physically motivated extension that preserves the successes of Starobinsky inflation while improving its fit to modern precision cosmological data.

\end{abstract}

\maketitle

\section{Introduction}

The Starobinsky model of inflation \cite{Starobinsky:1980te}, formulated within the framework of $f(R) = R + R^2/(6M^2)$ gravity, stands as a cornerstone of modern cosmological theory. Its predictions for a nearly scale-invariant spectrum of perturbations, characterized by the scalar spectral index $n_s$ and an exceptionally small tensor-to-scalar ratio $r$, have shown remarkable consistency with data from the \textit{Planck} satellite and \textit{BICEP/Keck} array, cementing its status as a benchmark model \cite{Ketov:2025nkr}. However, the advent of increasingly precise measurements from the \textit{Atacama Cosmology Telescope (ACT)} \cite{ACT:2025fju,ACT:2025tim,ACT:2025rvn} introduces nuanced tensions. The latest \textit{PACT-LB} dataset (combining \textit{Planck}, \textit{ACT}, and \textit{DESI} BAO data \cite{DESI:2024uvr}) reports a value of $n_s = 0.9743 \pm 0.0034$ \cite{ACT:2025fju}, which sits at the edge of the standard Starobinsky prediction and motivates the exploration of natural extensions to the theory. For a view of the ongoing discourse, see the recent literature in Refs.~\cite{Kallosh:2025rni,Aoki:2025wld,Berera:2025vsu,Brahma:2025dio,Dioguardi:2025vci,Gialamas:2025kef,Salvio:2025izr,Kim:2025dyi,Antoniadis:2025pfa,Dioguardi:2025mpp,Gao:2025onc,Pallis:2025epn,Drees:2025ngb,Zharov:2025evb,Haque:2025uri,Liu:2025qca,Cheng:2025lod,Yin:2025rrs,Gialamas:2025ofz,Haque:2025uis,Addazi:2025qra,Peng:2025bws,Frolovsky:2025iao,Pallis:2025nrv,Wang:2025dbj,Wolf:2025ecy,McDonald:2025tfp,Choudhury:2025vso,Gao:2025viy,Pallis:2025gii,SidikRisdianto:2025qvk,Zahoor:2025nuq,Ye:2025idn,Ketov:2025cqg,Zhu:2025twm,Ellis:2025ieh,Ivanov:2025nsx,Yuennan:2025kde,Modak:2025bjv,Choi:2025qot,Odintsov:2025eiv,Aoki:2025ywt,Pallis:2025vxo,Aldabergenov:2025kcv,Yogesh:2025wak,Mohammadi:2025gbu,Kouniatalis:2025orn}.

In this work, we investigate a specific and well-motivated extension of $f(R)$ gravity,~\footnote{For foundational reviews on $f(R)$ and modified gravity, see Refs.~\cite{Capozziello:2007ec,Capozziello:2011et,DeFelice:2010aj,Nojiri:2006gh}.} based on a general function $f(L)$ of the 4D Lovelock invariant $L = -2\lambda + R + \frac{\alpha}{4}\cg$, consisting of a constant term $-2\lambda$, scalar curvature $R$, and the Gauss--Bonnet (GB) term
\begin{equation}
    \cg\equiv R^2-4R_{\mu\nu}R^{\mu\nu}+R_{\mu\nu\rho\sigma}R^{\mu\nu\rho\sigma}.
\end{equation}
This type of modified gravity was studied in Ref. \cite{Bueno:2016dol} in general spacetime dimensions. Here we consider its application to inflation by Taylor-expanding $f(L)$ up to the quadratic term, $f(L) = L + L^2/(6M^2)$, in analogy with Starobinsky gravity. This model is not merely another higher-derivative correction; it represents a specific class of theories where the scalar-tensor dual, after a Weyl transformation, features a scalar field with non-minimal coupling to the Gauss-Bonnet term and a specific higher-derivative interactions which can be shown to be a particular case of Horndeski gravity \cite{Horndeski:1974wa}. The latter is known to produce second-order equations of motion, avoiding ghosts associated with higher time derivatives. At the same time, the new non-minimal interactions, absent in the pure $f(R)$ case, introduce distinctive modifications to the inflationary dynamics. Furthermore, we find that the quadratic model with $f(L) = L + L^2/(6M^2)$, is equivalent to Higgs inflation coupled to the Gauss--Bonnet term in the Jordan frame.

We demonstrate that the GB coupling $\alpha$ in this $f(L)$ framework provides a novel mechanism to reconcile the inflationary predictions with the \textit{ACT} data. The derivative couplings inherent to the model's scalar-tensor representation alter the slow-roll parameters in a way that can increase the predicted value of $n_s$, bringing it into closer alignment with the ACT measurement, while simultaneously offering a potentially testable signal in the tensor-to-scalar ratio $r$. Our findings position this $f(L)$ gravity model as a compelling and theoretically coherent refinement of the Starobinsky paradigm, capable of addressing emerging observational nuances without abandoning its foundational successes.

\section{The Model}

We begin with a modified theory of gravity, so-called 4D $f({\rm Lovelock})$ gravity, defined by the Lagrangian density (we assume Planck units, $M_P=1$),
\begin{equation}
\label{Lov1}
\mathcal{L}=\frac{1}{2}\sqrt{-g}~f(L)~,~~~L=-2\lambda+R+\frac{\alpha}{4}\cg~,
\end{equation}
where $L$ is the 4d Lovelock combination with a ``cosmological constant" term $\lambda$ and a GB coupling parameter $\alpha$. An equivalent form of this Lagrangian is given by
\begin{equation}
\label{Lov2}
\sqrt{-g}^{-1}\mathcal{L}=\frac{1}{2}[f'(Z)L-f'(Z)Z+f(Z)]~,
\end{equation}
where varying w.r.t. the auxiliary scalar $Z$ yields $Z=L$, leading to the Lagrangian \eqref{Lov1}, provided that $f''(Z)\neq 0$. As can be seen, one scalar degree of freedom is enough to describe scalar-tensor formulation of this theory (see below), in contrast to general $f(R,\cg)$ gravities which require two scalars, one of which is often a ghost \cite{DeFelice:2010hg} (see also \cite{Bamba:2010wfw,Elizalde:2010jx,Atazadeh:2013cz,Nojiri:2021mxf}).

The Lagrangian \eqref{Lov2} is transformed from the Jordan frame to the Einstein frame via a Weyl rotation:
\begin{equation}
\label{Conformal}
g_{\mu\nu}\rightarrow \frac{1}{f'}g_{\mu\nu}~.
\end{equation}
Here $f$ should be understood as a function of $Z$. After the transformation \eqref{Conformal}, the Lagrangian (up to total derivatives) becomes,
\begin{align}
\begin{aligned}
\label{LL}
\frac{\mathcal{L}}{\sqrt{-g}} &=\frac{1}{2}R-\frac{3}{4f'^{2}}\partial f' \partial f'-\frac{1}{2f'}\Big(Z+2\lambda-\frac{Z}{f'}\Big)\\
&+\frac{\alpha}{8}\Big[f'\cg+\frac{4}{f'}G^{\mu\nu}\partial_{\mu} f' \partial_{\nu} f'\\
&\qquad\quad-\frac{3}{f'^2}\partial f' \partial f'\Box f'+\frac{3}{f'^{3}}(\partial f' \partial f')^{2}\Big]~,
\end{aligned}
\end{align}
where $\Box\equiv\nabla_\mu\nabla^\mu$ ($\nabla_\mu$ is the covariant derivative), and $G_{\mu\nu}=R_{\mu\nu}-\frac{1}{2}g_{\mu\nu}R$ is the Einstein tensor. The derivative $f'$ is then promoted to a canonical scalar field $\varphi$ via the redefinition,
\begin{equation}
\label{Transf}
f'=e^{\sqrt{\frac{2}{3}}\varphi}~.
\end{equation}
Substituting this definition into the transformed Lagrangian \eqref{LL} leads to the final form in the Einstein frame:
\begin{align}
\begin{aligned}
\label{L_canonical}
&\sqrt{-g}^{-1}\mathcal{L}=\frac{1}{2}R-\frac{1}{2}\partial \varphi \partial \varphi -V(\varphi)\\
&\quad +\frac{\alpha}{8} e^{\sqrt{\frac{2}{3}}\varphi}\Big(\cg+\frac{8}{3}G^{\mu\nu}\partial_{\mu}\varphi\partial_{\nu}\varphi-\sqrt{\frac{8}{3}}\partial \varphi \partial \varphi \Box \varphi\Big)~,
\end{aligned}
\end{align}
with the scalar potential
\begin{equation}
    V(\varphi)=\tfrac{1}{2}e^{-\sqrt{\frac{2}{3}}\varphi}\big[Z(\varphi)+2\lambda-e^{-\sqrt{\frac{2}{3}}}f(Z(\varphi))\big]~,
\end{equation}
where $Z(\varphi)$ is to be found from Eq. \eqref{Transf}. The Lagrangian \eqref{L_canonical} is a special case of Horndeski gravity \cite{Horndeski:1974wa}, a.k.a. generalized galileons \cite{Deffayet:2011gz}, which is the most general scalar-tensor gravity having second-order equations of motion (see, e.g., \cite{Kobayashi:2019hrl} for a review of Horndeski theory, and \cite{Meissner:1996sa,Easson:2020bgk} for its connection to string theory).

To provide a concrete example, we specify the function $f(L)$ as an expansion up to the second order:
\begin{equation}
\label{tra}
f(L)=L+\frac{L^2}{6M^{2}}~. 
\end{equation}
For this model, the relations defining the scalar sector from Eq. \eqref{Lov2} are:
\begin{gather}
\begin{gathered}
f(Z)=Z+\frac{Z^{2}}{6M^{2}}~,~~~f'(Z)=e^{\sqrt{\frac{2}{3}}\varphi}=1+\frac{Z}{3M^{2}}~,\\
\Rightarrow~~Z(\varphi)=3M^{2}\Big(e^{\sqrt{\frac{2}{3}}\varphi}-1\Big)~,
\end{gathered}
\end{gather}
The scalar potential is derived as
\begin{equation}
\label{potV}
V(\varphi)=\tfrac{3}{4}M^{2}\Big(1-e^{-\sqrt{\frac{2}{3}}\varphi}\Big)^{2}+\lambda e^{-\sqrt{\frac{2}{3}}}~,
\end{equation}
which is exactly the Starobinsky potential of $R^2$ gravity plus a $\lambda$-term, which can shift the Minkowski vacuum ($\lambda=0$) to de Sitter if $\lambda>0$. For small enough $\lambda$, the resulting cosmological constant can describe dark energy, in which case $\lambda$ can be ignored during inflation, so we set it to zero in the rest of the paper. The parameter $M$ is the scalaron/inflaton mass which is fixed as $M\sim 10^{-5}$ by the observed amplitude of scalar perturbations.

It follows that the model \eqref{L_canonical} with quadratic $f(Z)$ is a one parameter extension of the Starobinsky model (for $\lambda=0$) by a set of Horndeski-type higher derivative terms, including a GB non-minimal coupling. Furthermore, it is interesting to note that the resulting theory is equivalent to the model of Higgs inflation coupled to the GB term in the Jordan frame, studied in Ref. \cite{Koh:2023zgn} (see also \cite{Guleryuz:2026jam} for a GB-corrected Starobinsky model). More specifically, the action \eqref{L_canonical} with the potential \eqref{potV} coincides with Eq. (2.18) of \cite{Koh:2023zgn}, which describes the Einstein frame action of GB-coupled Higgs inflation (where the GB coupling and the scalar curvature coupling are proportional to each other) in the large field limit. This equivalence generalizes the well-known equivalence between Starobinsky inflation and Higgs inflation.

Let us now derive the equations of motion following from \eqref{L_canonical}. The Klein--Gordon equation and the Einstein equations are given by
\begin{widetext}
\begin{align}
\begin{split}\label{KG}
    &\Box \varphi- V_{,\varphi}-\tfrac{1}{8}\xi_{,\varphi}\Big(\cg+\tfrac{8}{3}G^{\mu\nu}\partial_\mu\varphi\partial_\nu\varphi-\sqrt{\tfrac{8}{3}}\partial\varphi\partial\varphi\Box\varphi\Big)+\tfrac{1}{\sqrt{6}}\partial_\mu\xi\Big(\sqrt{\tfrac{8}{3}}G^{\mu\nu}\partial_\nu\varphi-\partial^\mu\varphi\Box\varphi+2\nabla^\mu\nabla^\nu\varphi\partial_\nu\varphi\Big)\\
    &+\tfrac{1}{2\sqrt{6}}\Box\xi\partial\varphi\partial\varphi+\tfrac{1}{\sqrt{6}}\xi\Big(\sqrt{\tfrac{8}{3}}G^{\mu\nu}\nabla_\mu\partial_\nu\varphi+R^{\mu\nu}\partial_\mu\varphi\partial_\nu\varphi-\Box\varphi\Box\varphi+\nabla^\mu\nabla^\nu\varphi\nabla_\mu\nabla_\nu\varphi\Big)=0~,
\end{split}\\[10pt]
\begin{split}\label{EFE}
    &(1+\Box\xi)G_{\mu\nu}-\partial_\mu\varphi\partial_\nu\varphi+\tfrac{1}{2}g_{\mu\nu}(\partial\varphi\partial\varphi+2V)+\tfrac{1}{2}\nabla_\mu\nabla_\nu\xi\,R+\nabla^\rho\nabla^\sigma\xi\,R_{\mu\rho\sigma\nu}+g_{\mu\nu}\nabla^\rho\nabla^\sigma\xi\,R_{\rho\sigma}\\
    &-\nabla^\rho\nabla_\mu\xi\,R_{\rho\nu}-\nabla^\rho\nabla_\nu\xi\,R_{\rho\mu}+\tfrac{1}{3}\nabla_\lambda\nabla_\mu(\xi\partial_\nu\varphi\partial^\lambda\varphi)+\tfrac{1}{3}\nabla_\lambda\nabla_\nu(\xi\partial_\mu\varphi\partial^\lambda\varphi)-\tfrac{1}{3}\nabla_\mu\nabla_\nu(\xi\partial\varphi\partial\varphi)-\tfrac{1}{3}\Box(\xi\partial_\mu\varphi\partial_\nu\varphi)\\
    &+\tfrac{1}{3}g_{\mu\nu}\Box(\xi\partial\varphi\partial\varphi)-\tfrac{1}{3}g_{\mu\nu}\nabla^\rho\nabla^\sigma(\xi\partial_\rho\varphi\partial_\sigma\varphi)-\tfrac{1}{2\sqrt{6}}\big[\nabla_\mu(\xi\partial\varphi\partial\varphi)\partial_\nu\varphi+\nabla_\nu(\xi\partial\varphi\partial\varphi)\partial_\mu\varphi-g_{\mu\nu}\nabla_\lambda(\xi\partial\varphi\partial\varphi)\partial^\lambda\varphi\big]\\
    &-\tfrac{1}{3}\xi\Big(2R_{\nu\lambda}\partial_\mu\varphi\partial^\lambda\varphi+2R_{\mu\lambda}\partial_\nu\varphi\partial^\lambda\varphi-g_{\mu\nu}G^{\rho\sigma}\partial_\rho\varphi\partial_\sigma\varphi-R\partial_\mu\varphi\partial_\nu\varphi-R_{\mu\nu}\partial\varphi\partial\varphi-\sqrt{\tfrac{3}{2}}\partial_\mu\varphi\partial_\nu\varphi\Box\varphi\Big)=0~,
\end{split}
\end{align}
\end{widetext}
where $\xi=-\alpha e^{\sqrt{2/3}\varphi}$.

We conclude this section by highlighting several key findings:

\begin{enumerate}
    \item Although \( f(L) \) is a subset of \( f(R,\cg) \) gravity, it is ghost-free, unlike the general \( f(R,\cg) \) case which typically contains ghost modes.

    \item The theory defined by $f(L)=L+L^2/(6M^2)$ is equivalent to Higgs--Gauss--Bonnet inflation \cite{Koh:2023zgn}, as seen from Eq. \eqref{L_canonical}. Additionally, its higher-derivative sector belongs to the generalized galileon/Horndeski class \cite{Deffayet:2013lga} (for example, from \eqref{EFE} one can show that all higher derivatives of $\varphi$ cancel out, and the equations are second-order). These non-trivial dualities are, to our knowledge, novel.
\end{enumerate}

\section{Inflationary solutions}

In the FLRW background $g_{\mu\nu}={\rm diag}(-1,a^2,a^2,a^2)$, the Klein--Gordon equation \eqref{KG} takes the form
\begin{align}\label{KG_FLRW}
\begin{aligned}
    \ddot\varphi &+3H\dot\varphi+V_{,\varphi}-\tfrac{1}{2\sqrt{6}}\ddot\xi\dot\varphi^2-\tfrac{1}{\sqrt{6}}\dot\xi\dot\varphi(2\sqrt{6}H^2+\ddot\varphi-\tfrac{3}{2}H\dot\varphi)\\
    &+\xi_{,\varphi}\Big[3H^2(\dot H+H^2)+\Big(H^2-\tfrac{1}{2\sqrt{6}}\ddot\varphi-\sqrt{\tfrac{3}{8}}H\dot\varphi\Big)\dot\varphi^2\Big]\\
    & -\tfrac{1}{\sqrt{6}}\xi\Big[2(\sqrt{6}H-3\dot\varphi)H\ddot\varphi+2\sqrt{6}(3H^2+2\dot H)H\dot\varphi\\
    &\hspace{4.2cm}-3(3H^2+\dot H)\dot\varphi^2\Big]~,
\end{aligned}
\end{align}
while the Einstein equations yield the Friedmann equations:
\begin{align}
\begin{split}\label{Friedmann1}
    &3(1-\dot\xi H)H^2-V\\
    &-\tfrac{1}{2}\Big(1-\tfrac{1}{\sqrt{6}}\dot\xi\dot\varphi-6\xi H^2+\sqrt{6}\xi H\dot\varphi\Big)\dot\varphi^2=0~,
\end{split}\\[10pt]
\begin{split}\label{Friedmann2}
    &2(1-\dot\xi H+\tfrac{1}{3}\xi\dot\varphi^2)\dot H-(\ddot\xi-\dot\xi H)H^2\\
    &+\tfrac{1}{3}\xi\Big(4H-\sqrt{\tfrac{3}{2}}\dot\varphi\Big)\ddot\varphi\dot\varphi+\Big(1+\tfrac{2}{3}\dot\xi H-\tfrac{1}{\sqrt{6}}\dot\xi\dot\varphi\\
    &\hspace{2.8cm}-2\xi H^2+\sqrt{\tfrac{3}{2}}\xi H\dot\varphi\Big)\dot\varphi^2=0~.
\end{split}
\end{align}

In order to describe slow-roll inflation, it is convenient to introduce a set of slow-roll parameters,
\begin{gather}
\begin{gathered}
    \epsilon\equiv -\frac{\dot H}{H^2}~,~~~\eta\equiv\frac{\ddot\varphi}{H\dot\varphi}~,\\
    \delta\equiv \xi\dot\varphi^2~,~~~\omega\equiv \dot\xi H~,~~~\sigma\equiv\frac{\dot\omega}{H\omega}~,
\end{gathered}
\end{gather}
so that slow-roll is defined by $\{|\epsilon|,|\eta|,|\delta|,|\omega|,|\sigma|\}\ll 1$.

\subsection{Slow-roll approximation}

We will first derive analytical slow-roll solutions under perturbative expansion in small GB parameter $|\alpha|$. Therefore, these solutions are perturbations around the pure Starobinsky/Higgs inflation, which will then be compared to full numerical solutions providing more accurate results.

Under the slow-roll conditions $\{|\epsilon|,|\eta|,|\delta|,|\omega|,|\sigma|\}\ll 1$, the equations of motion \eqref{KG_FLRW}, \eqref{Friedmann1}, and \eqref{Friedmann2} reduce to
\begin{align}
    3H\dot\varphi(1-2\xi H^2) &\simeq -V_{,\varphi}-3\xi_{,\varphi}H^4~,\label{KG_approx}\\
    3H^2 &\simeq V~,\label{F1_approx}\\
    \epsilon-\tfrac{1}{2}\omega+\delta &\simeq\dot\varphi^2/(2H^2)~.\label{F2_approx}
\end{align}
From \eqref{KG_approx} one can read off the effective potential slope
\begin{equation}
    V^{\rm eff}_{,\varphi}=\frac{V_{,\varphi}+\tfrac{1}{3}\xi_{,\varphi}V^2}{1-\tfrac{2}{3}\xi V}~,
\end{equation}
where we used $3H^2\simeq V$, and $V$ is the Starobinsky potential $V=\tfrac{3}{4}M^2(1-e^{-\sqrt{2/3}\varphi})^2$. It is convenient to switch from physical time $t$, to the (forward) number of e-folds satisfying $\dot N=H$. Equation \eqref{KG_approx} can then be written as
\begin{equation}\label{KG_approx2}
    \varphi'(N)\simeq -V^{\rm eff}_{,\varphi}/V~,
\end{equation}
where $'\equiv d/dN$. By using the notation $y\equiv e^{-\sqrt{2/3}\varphi}$ and $\hat\alpha\equiv\alpha M^2$, we write \eqref{KG_approx2} as
\begin{equation}\label{KG_y_approx}
    \frac{y'}{y}\simeq \frac{4y^2-\tfrac{1}{2}\hat\alpha(1-y)^3}{3(1-y)y+\tfrac{3}{2}\hat\alpha(1-y)^3}~.
\end{equation}
Assuming $y\ll 1$ (since inflation requires $\varphi\gg 1$) and $|\hat\alpha|/y^2\ll 1$, perturbative solution to  \eqref{KG_y_approx} can be found as
\begin{equation}\label{N_of_y}
    N(y)\simeq\frac{3}{4}\Big(\frac{1}{y_*}-\frac{1}{y}\Big)+\frac{\hat\alpha}{32}\Big(\frac{1}{y_*^3}-\frac{1}{y^3}\Big)~,
\end{equation}
where we use the convention that $N=0$ at the horizon exit of the CMB reference scale $k_*$ (e.g., $0.05~{\rm Mpc}^{-1}$), and $y_*$ is the corresponding inflaton value. The value $y_*$ can be estimated by using the fact that $y_e\gg y_*$, where subscript `e' denotes the value at the end of inflation. Thus, from \eqref{N_of_y} we have
\begin{equation}\label{y_pert_sol}
    N_e\simeq \frac{3}{4y_*}+\frac{\hat\alpha}{32y_*^3}~~\Rightarrow~~y_*\simeq \frac{3}{4N_e}+\frac{\hat\alpha N_e}{18}~,
\end{equation}
where we assume $\hat\alpha N_e^2\ll 1$, and $N_e$ is the e-fold number at the end of inflation, which in our notation coincides with the total number of e-folds from the horizon exit (i.e., $50\lesssim N_e\lesssim 60$). With this we can determine the approximate value of $y_*$ for a given $\hat\alpha$, and estimate the inflationary observables $n_s$ and $r$.

\subsection{Estimating $n_s$ and $r$}

Scalar spectral tilt $n_s$ and tensor-to-scalar ratio $r$ for the model \eqref{L_canonical} are given by \cite{Koh:2023zgn,Hwang:2005hb}~\footnote{In \cite{Hwang:2005hb} inflationary perturbations were studied in a string-inspired class of modified gravities. Our Lagrangian \eqref{L_canonical}, equivalent to the Higgs--Gauss--Bonnet model of \cite{Koh:2023zgn}, belongs to this class.}
\begin{gather}
    n_s\simeq 1-4\epsilon-2\eta-2q~,\label{n_s_eq}\\[10pt]
    r\simeq\bigg|\frac{16C_s^3/C_t^3}{1-\omega+\tfrac{1}{3}\delta}\Big(\frac{{\varphi'}^2}{2}-\frac{\omega{\varphi'}^3}{\sqrt{6}}-\delta+\sqrt{\tfrac{3}{2}}\delta\varphi'+\omega\sigma\Big)\bigg|~,\label{r_eq}
\end{gather}
where
\begin{widetext}
\begin{align}
\begin{split}
    q &\equiv \Bigg[2{\varphi'}^2-4\delta+2\sqrt{6}\delta\varphi'-\sqrt{\tfrac{8}{3}}\omega{\varphi'}^3+\frac{3(\omega-\tfrac{4}{3}\delta+\tfrac{1}{\sqrt{6}}\delta\varphi')^2}{1-\omega+\tfrac{1}{3}\delta}\Bigg]^{-1}\Bigg\{4\delta(\epsilon+\eta)-2\delta'-\sqrt{6}\delta\Big(\epsilon+\eta-\frac{\delta'}{\delta}\Big)\varphi'\\
    &\qquad-\sqrt{\frac{2}{3}}\omega(\epsilon+\eta+\sigma){\varphi'}^3-\frac{3(\omega-\tfrac{4}{3}\delta+\tfrac{1}{\sqrt{6}}\delta\varphi')}{1-\omega+\tfrac{1}{3}\delta}\Big[\omega(\epsilon-\sigma)-\tfrac{4}{3}(\epsilon\delta-\delta')-\tfrac{1}{\sqrt{6}}(\eta\delta+\delta')\varphi'\Big]\\
    &\qquad-\frac{3(\omega-\tfrac{4}{3}\delta+\tfrac{1}{\sqrt{6}}\delta\varphi')^2}{2(1-\omega+\tfrac{1}{3}\delta)^2}\big[2\eta(1-\omega+\tfrac{1}{3}\delta)-\omega\sigma+\tfrac{1}{3}\delta'\big]\Bigg\}~,
\end{split}\\[10pt]
\begin{split}
    C_s^2 &\equiv1+\Bigg[{\varphi'}^2-2\delta+\sqrt{6}\delta\varphi'-\sqrt{\tfrac{2}{3}}\omega{\varphi'}^3+\frac{3(\omega-\tfrac{4}{3}\delta+\tfrac{1}{\sqrt{6}}\delta\varphi')^2}{2(1-\omega+\tfrac{1}{3}\delta)}\Bigg]^{-1}\Bigg\{\tfrac{4}{3}\delta\epsilon-\sqrt{\tfrac{2}{3}}\delta(1-\eta)\varphi'+\sqrt{\tfrac{2}{3}}\omega{\varphi'}^3\\
    &\qquad-\frac{\omega-\tfrac{4}{3}\delta+\tfrac{1}{\sqrt{6}}\delta\varphi'}{1-\omega+\tfrac{1}{3}\delta}\Big[2\omega\epsilon-\tfrac{4}{3}\delta(1-\eta)+\sqrt{\tfrac{2}{3}}\delta\varphi'+\tfrac{2}{3}\omega{\varphi'}^2\Big]+\frac{(\omega-\tfrac{4}{3}\delta+\tfrac{1}{\sqrt{6}}\delta\varphi')^2}{2(1-\omega+\tfrac{1}{3}\delta)^2}\big[\tfrac{2}{3}\delta-\omega(1-\epsilon-\sigma)\big]\Bigg\}~,
\end{split}\\[10pt]
\begin{split}
    C_t^2 &\equiv\frac{1-\tfrac{1}{3}\delta-\omega(\epsilon+\sigma)}{1-\omega+\tfrac{1}{3}\delta}~.
\end{split}
\end{align}
\end{widetext}
One can greatly simplify these expressions by using the slow-roll perturbative solution \eqref{y_pert_sol}. The slow-roll parameters for this solution, at the horizon exit and linear order at $\hat\alpha$, are given by
\begin{gather}
\begin{gathered}
    \epsilon\simeq \frac{3}{4N_e^2}-\frac{\hat\alpha}{18}~,~~~\eta\simeq\frac{1}{N_e}+\frac{8}{27}\hat\alpha N_e~,\\
    \delta\simeq -\frac{\hat\alpha}{2N_e}~,~~~\omega\simeq\frac{\hat\alpha}{3}~,~~~\sigma\simeq -\frac{3}{2N_e^2}+\frac{\hat\alpha}{9}~.
\end{gathered}
\end{gather}
The observables \eqref{n_s_eq} and \eqref{r_eq} then reduce to
\begin{align}
    n_s &\simeq 1-\frac{2}{N_e}\Big(1+\frac{8}{27}\hat\alpha N_e^2+\co(\hat\alpha^2 N_e^4)\Big)~,\label{n_s_approx}\\[10pt]
    r &\simeq \frac{12}{N_e^2}\Big(1-\frac{8}{27}\hat\alpha N_e^2+\co(\hat\alpha^2 N_e^4)\Big)~.\label{r_approx}
\end{align}
The predictions of the baseline Starobinsky model are recovered in the case $\hat\alpha=0$. Significant deviations from it occur when $|\hat\alpha|$ approaches $10^{-3}$ from below (given that $N_e\sim 50-60$), at which point the perturbative solution breaks down. It can also be seen that negative $\hat\alpha$ is needed in order to increase the value of $n_s$ for a better alignment with the ACT data. This, in turn, will increase the value of $r$, which could be probed by next generation of CMB experiments.

\begin{figure*}
 \includegraphics[width=.9\textwidth]{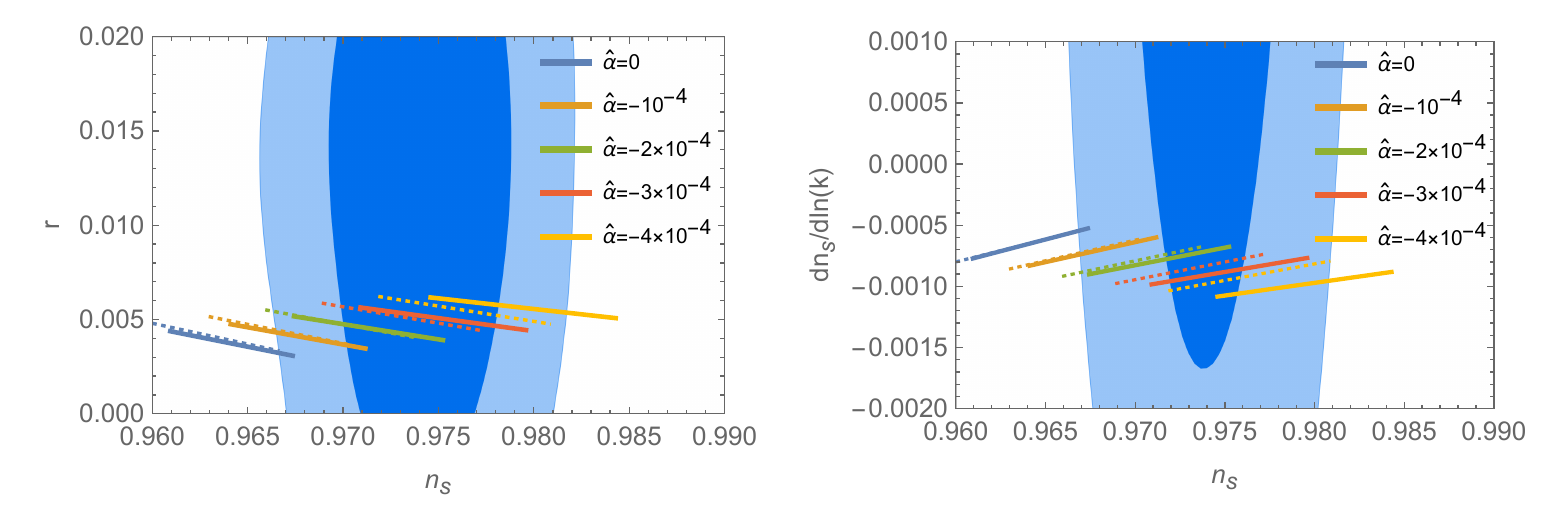}
\caption{Predictions of $({\rm Lovelock})^2$ inflation for the spectral index $n_s$, tensor-to-scalar ratio $r$, and the running $dn_s/d\ln k=-dn_s/dN$, compared to the Planck+ACT constraints \cite{ACT:2025tim}. Solid lines represent the results of numerical integration of the background equations of motion, and the dashed lines (of the same color) represent the corresponding approximations \eqref{n_s_approx} and \eqref{r_approx}. The calculations are done for $50\leq N_e\leq 60$ (larger $N_e$ leads to larger $n_s$).}
\label{Fig_ns_r}
\end{figure*}

The plots of the approximated $n_s$ \eqref{n_s_approx} and $r$ \eqref{r_approx} are shown in Fig. \ref{Fig_ns_r}, compared to the results from numerical integration of the background equations \eqref{KG_FLRW}, \eqref{Friedmann1}, and \eqref{Friedmann2}. It can be seen that the approximations start to noticeably deviate from the numerical results for larger $|\hat\alpha|$, approaching $10^{-3}$, as expected. Figure \ref{Fig_ns_r} implies that for a better fit with Planck+ACT constraints, $\hat\alpha$ should be close to $-3\times 10^{-4}$.

As for the amplitude $A_s$ of scalar perturbations, during slow-roll it can be approximated as \cite{Hwang:2005hb}
\begin{align}
\begin{aligned}
    A_s\simeq \frac{H^2}{4\pi^2{\varphi'}^2} &\simeq\frac{V}{12\pi^2(2\epsilon-\omega+2\delta)}\\
    &\simeq \frac{M^2N_e^2}{24\pi^2(1-\tfrac{8}{27}\hat\alpha N_e^2)}~,
\end{aligned}
\end{align}
where we used the slow-roll solution \eqref{y_pert_sol}. For the Planck normalization $A_s\approx 2.1\times 10^{-9}$, and $N_e=55$ e-folds, we get
\begin{equation}
    1.28\times 10^{-5}\lesssim M\lesssim 1.5\times 10^{-5}~,
\end{equation}
for the values $0\geq\hat\alpha\geq -4\times 10^{-4}$ considered in Fig. \ref{Fig_ns_r}.

\subsection{Perturbative unitarity: estimate of the cutoff scale}

We now provide a quantitative estimate of the cutoff scale $\Lambda_U$ in the Einstein frame, following the standard power-counting arguments (see e.g. Refs.~\cite{Burgess:2009ea,Barbon:2009ya,Lerner:2009na,Burgess:2010zq,Hertzberg:2010dc,Bezrukov:2010jz,Calmet:2013hia,Kehagias:2013mya,Escriva:2016cwl,Antoniadis:2021axu,Ito:2021ssc,He:2026fzs}).

The dimensionful GB coupling $\alpha$ is related to $\hat\alpha$ by $\alpha = \hat\alpha / M^2$, with $M \sim 1.4 \times 10^{-5}$ (in Planck units) and $\hat\alpha \sim -3\times 10^{-4}$ (the preferred value to fit the ACT data). This gives
\begin{equation}
    |\alpha| \sim \frac{3\times 10^{-4}}{(1.4\times 10^{-5})^2} \sim 1.5 \times 10^{6}~.
\end{equation}
Restoring $M_P$, $|\alpha| \sim 1.5\times 10^{6} M_P^{-2}$. The associated GB mass scale is
\begin{equation}
    M_{\mathcal{G}} \equiv |\alpha|^{-1/2} M_P \sim \frac{M_P}{\sqrt{1.5\times 10^{6}}} \sim 2\times 10^{15} \text{ GeV}~.
\end{equation}
This is the scale that controls the higher-derivative interactions.

\textbf{Cutoff estimate:}
Here we will use the Einstein-frame action \eqref{L_canonical}, and focus on the new interactions (compared to the Starobinsky model) comprising the three terms in the second line of \eqref{L_canonical}. By expanding around the background field $\bar\varphi$ and flat metric, and introducing perturbations $\hat\varphi$ (inflaton) and $h_{\mu\nu}$ (graviton) of mass dimension one, the aforementioned three Lagrangian terms lead to the new vertex types proportional to $\alpha$: $h-h-h$, $h-h-\hat\varphi$, $h-\hat\varphi-\hat\varphi$, and $\hat\varphi-\hat\varphi-\hat\varphi$. From these one can construct tree-level scattering diagrams $hh\rightarrow hh$, $h\hat\varphi\rightarrow h\hat\varphi$, and $\hat\varphi\hat\varphi\rightarrow \hat\varphi\hat\varphi$ with either graviton or inflaton propagator. Power-counting estimate of the resulting scattering amplitudes leads to the cutoff scale~\footnote{One can also introduce a dimensionless parameter $\gamma\equiv M_P^2\hat\alpha/(M^2\bar y)$, such that the leading GB-scalar coupling term (in the expansion around the inflationary background) is $\sim \gamma M_P^{-1}\hat\varphi\cg$. Then the cutoff scale is $\Lambda_U\sim M_P|\gamma|^{-1/3}$.}
\begin{equation}\label{cut_off_Lovelock}
    \Lambda_U\sim\bigg(\frac{M^2\bar y}{M_P^2|\hat\alpha|}\bigg)^{1/3}\times M_P\sim 10^{-3} M_P~.
\end{equation}
Here, $\bar y\equiv e^{-\sqrt{2/3}\bar\varphi}$ is the inflaton background value estimated using \eqref{y_pert_sol} (we take $N_e=55$ e-folds), $M\approx 1.4\times 10^{-5}M_P$ is the inflaton mass parameter, and $\hat\alpha=-3\times 10^{-4}$ is the dimensionless GB parameter for the best fit to the ACT data. The estimate \eqref{cut_off_Lovelock} is universal for the three new terms ($\sim\cg$, $G^{\mu\nu}\partial_\mu\varphi\partial_\nu\varphi$, and $\partial\varphi\partial\varphi\Box\varphi$) of Eq. \eqref{L_canonical}.

The cutoff scale \eqref{cut_off_Lovelock} is larger than the inflationary Hubble scale, but of the same order as the inflationary energy density $V^{1/4}_{\rm inf}\sim \sqrt{MM_P}\sim 10^{-3}M_P$, raising concerns regarding the validity of classical approximation. We recall, however, that the same power-counting argument in the pure Starobinsky model leads to the cutoff $\Lambda_U\sim M_P\zeta^{-1/3}\sim 10^{-3}M_P$ \cite{Burgess:2009ea} (here $\zeta\equiv\frac{M_P^2}{12M^2}$), which is of the same order as \eqref{cut_off_Lovelock}. Nontheless, a more careful analysis of \cite{Hertzberg:2010dc} later revealed that due to cancellations between tree-level diagrams, the actual cutoff scale of the Starobinsky model is the Planck mass $M_P$. Therefore, this motivates a more detailed analysis of the scattering processes in our model (and in GB-coupled models in general), to see if such cancellations can occur here as well.

\subsection{Impact of the new Lovelock terms on reheating}

Due to the functional form of the GB coupling, $\sim e^{\sqrt{2/3}\varphi}$, the GB term has the strongest impact on the field dynamics during early inflation, when $\varphi\gg 1$. In particular, we showed that it can raise the scalar tilt $n_s$ compared to the pure Starobinsky model. As $\varphi$ decreases towards the end of inflation, the GB term becomes less important. And at the reheating stage, its effects become negligible, as will shown now.

Approximating $V(\varphi)$ around the bottom of the potential as $M^2\varphi^2/2$, the usual oscillatory solution for $\varphi$ during reheating is $\varphi\simeq \Phi\sin(Mt)$, where $\Phi$ is a slowly decaying amplitude. At the same time, the Hubble function evolves as $H\simeq M^2\Phi^2/\sqrt{6}$. To see that this solution holds in our model, we look at the Klein--Gordon Eq. \eqref{KG_FLRW} and Friedmann Eq. \eqref{Friedmann1} with $\xi=-\alpha e^{\sqrt{2/3}\varphi}$. Let us split \eqref{KG_FLRW} as $\ca_{\rm Star}+\ca_{\rm GB}=0$, where $\ca_{\rm Star}$ is the Starobinsky part (first three terms of \eqref{KG_FLRW}), and $\ca_{\rm GB}$ include the terms proportional to $\alpha\equiv\hat\alpha/M^2$. Similarly, \eqref{Friedmann1} is split as $\cb_{\rm Star}+\cb_{\rm GB}=0$, where $\cb_{\rm Star}$ includes terms without $\alpha$, and $\cb_{\rm GB}$ includes terms with $\alpha$. By using $\varphi\sim \co(\Phi)$, $H\sim\dot\varphi\sim\co(M\Phi)$, and $\ddot\varphi\sim\co(M^2\Phi)$, one can show that 
\begin{gather}
\begin{gathered}
    \ca_{\rm Star}\sim \co(10^{-10}\Phi)~,~~~\ca_{\rm GB}\sim\co(10^{-14}\Phi^3)~,\\
    \cb_{\rm Star}\sim\co(10^{-10}\Phi^2)~,~~~\cb_{\rm GB}\sim\co(10^{-14}\Phi^4)~,
\end{gathered}
\end{gather}
where we took $M\sim 10^{-5}$ and $\hat\alpha\sim 10^{-4}$ (all in Planck units). It can be seen that during reheating there is a huge suppression of the GB and non-linear derivative terms, as $\ca_{\rm GB}/\ca_{\rm Star}\sim\co(10^{-4}\Phi^2)$ and $\cb_{\rm GB}/\cb_{\rm Star}\sim\co(10^{-4}\Phi^2)$, noting also that $\Phi$ is smaller than one already at the start of reheating. Therefore, the reheating dynamics is dictated by the usual Starobinsky terms, justifying the standard approximation $\varphi\simeq \Phi\sin(Mt)$ and $H\simeq M^2\Phi^2/\sqrt{6}$, where the amplitude decays as $\Phi\propto 1/(Mt)$.

The reheating (and preheating) can proceed according to the standard lore \cite{Kofman:1997yn}, by coupling $\varphi$ to matter. The reheating temperature can be estimated as $T_{\rm reh}\simeq 0.2\sqrt{\Gamma M_P}$, where $\Gamma$ is the decay rate of $\varphi$ into matter particles. Alternatively, purely gravitational reheating is also possible in the Starobinsky model, with $T_{\rm reh}\sim 10^8-10^9$ GeV from either minimally or non-minimally coupled matter \cite{Dorsch:2024nan,Dorsch:2026ref}.

\subsection{Radiative stability of $\hat\alpha$}

We now provide an explicit estimate of radiative corrections to the GB parameter $\alpha$. Loop corrections to $\alpha$ arise from diagrams involving the exponential coupling $\alpha e^{\sqrt{2/3}\varphi} \mathcal{G}$ (and from the non-linear $\varphi$-derivative terms, which leads to the same estimates below). Expanding around the inflationary background, the leading order term corresponds to 
$\alpha e^{\sqrt{2/3}\bar\varphi} \hat\varphi \mathcal{G}$ . At one loop, the correction scales as
\begin{equation}
    \delta \alpha \sim \frac{\Lambda^2}{16\pi^2 M_P^2} \alpha~,
\end{equation}
where $\Lambda$ is the cutoff. Using our estimated cutoff from Eq. \eqref{cut_off_Lovelock}, we get $\delta\alpha/\alpha\sim 10^{-8}$. Thus, radiative corrections to $\hat\alpha$ are suppressed at the level of $10^{-8}$. At the same time from Fig. \ref{Fig_ns_r} it is clear that sensitivity of the inflationary observables on precise value of $\alpha$ is mild -- as long as $\delta\alpha/\alpha\ll 1$, the predictions are radiatively stable.

\section{Conclusions}
\label{Sec_concl}

We have investigated the inflationary predictions of the ghost-free \( f(L) \) theory, where $L=R+\tfrac{\alpha}{4}\cg$ (up to a cosmological constant term which is irrelevant for inflation) is the 4D Lovelock invariant. We considered a simple quadratic form, \( f(L) = L + L^2/(6M^2) \), which is a one parameter extension of the Starobinsky gravity, with $M$ retaining its original meaning of the scalaron/inflaton mass. Motivated by the latest ACT observations, we derived the scalar-tensor formulation of the theory and computed slow-roll inflationary solution as a perturbative series in the GB parameter $\alpha$. This leads to particularly simple expressions for the spectral index $n_s$ and tensor-to-scalar ratio $r$, given by Eqs. \eqref{n_s_approx} and \eqref{r_approx}, which allows for a qualitative assessment of the deviations from the baseline Starobinsky model. The analytical results for $n_s$ and $r$ are plotted in Fig. \ref{Fig_ns_r} and compared to the results from numerical integration of the equations of motion. Our analysis demonstrates that this specific realization of the \( f(L) \) framework can produce values of $n_s$ and $r$ (and the running of $n_s$) that are in excellent agreement with current data.

The $f(L)$ gravity framework exhibits several remarkable properties that distinguish it from other modified gravity models. First, its scalar-tensor formulation, with a specific set of higher-derivative interactions, belongs to the Horndeski/galileon class, and guarantees the absence of Ostrogradsky instability. Second, the quadratic form $f=L+L^2/(6M^2)$ turns out to be equivalent to Higgs inflation coupled to the GB term (in the Jordan frame), extending the equivalence between the usual Starobinsky and Higgs inflationary scenarios.

The quadratic \( f(L) \) theory thus represents a compelling framework for addressing current observational tensions while maintaining theoretical consistency. Future work could explore more general forms of \( f(L) \) (including interactions with scalar fields), as well as investigate the reheating dynamics and primordial black hole production within this specific quadratic realization, particularly in light of its connections to fundamental ultraviolet completions of gravity.

Possible perturbative unitarity violation related to large non-minimal couplings may be of concern. Extensive literature exists on the non-minimal scalar curvature coupling of the Higgs field, as well as the Starobinsky model, see e.g. Refs. \cite{Burgess:2009ea,Barbon:2009ya,Lerner:2009na,Burgess:2010zq,Hertzberg:2010dc,Bezrukov:2010jz,Calmet:2013hia,Kehagias:2013mya,Escriva:2016cwl,Antoniadis:2021axu,Ito:2021ssc,He:2026fzs}. Following the standard power-counting arguments, we estimated the cutoff scale of our model to be $\Lambda_U\sim 10^{-3}M_P$ for the ACT-aligned GB parameter choice. This estimate is of the same order as the early result for the $R+R^2$ case \cite{Burgess:2009ea}. Notably, it was later shown in \cite{Hertzberg:2010dc} (and the subsequent works), that the actual cutoff scale of the Starobinsky model is lifted to the Planck scale, thanks to the cancellation between tree-level graviton scattering diagrams. This warrants a more detailed study of the scattering processes in the presence of the GB term, which can also be interesting for general inflationary models, where the GB term influences inflation.

When formulating the $L+L^2$ model as non-minimally coupled Higgs (or Higgs-like) theory, the magnitude of the scalar-$R$ coupling is at the same order as in the usual Higgs inflation scenarios~\footnote{Although such non-minimal coupling is often assumed to be large, when taking into account the top-quark mass, the effective Higgs potential at the inflationary scale can be modified to accommodate significantly smaller values of it, at the expense of larger tensor-to-scalar ratio -- see \cite{Hamada:2013mya,Hamada:2014iga,Hamada:2014wna}.}, while the scalar-GB coupling provides subleading or marginal corrections to the inflationary observables.

After inflation, as the scalaron oscillates around the minimum of its potential, we have shown that the GB and non-linear derivative terms are strongly suppressed. Thanks to this suppression, the process of reheating is determined, as usual, by the canonical kinetic term of the scalaron and its potential, as well as the coupling to matter fields (either direct coupling to $\varphi$ in the Einstein frame, or gravitational coupling in the Jordan frame).

As for the radiative stability of the GB parameter $\alpha$, the situation parallels Starobinsky inflation~\cite{Ellis:2025bzi}: loop corrections to $\alpha$ itself are suppressed to $\sim 10^{-8}$ for our estimated cutoff scale. However, as noted in Ref.~\cite{Bhattacharjee:2016ohe}, unitarity bounds and radiative corrections in scalar-GB theories depend sensitively on the functional form of the coupling.

{\bf Acknowledgements}. 
We thank the anonymous  reviewers for their constructive comments. 
AA work is supported by the Program for Innovative Research Team in Anqing Normal University and by the National Science Foundation of China (NSFC) through grant No. 12350410358.

\newpage

\providecommand{\href}[2]{#2}\begingroup\raggedright\endgroup

\end{document}